\documentclass{article}
\usepackage[utf8]{inputenc}

\usepackage{geometry}
\usepackage{amsmath}
\usepackage{amsfonts}
\usepackage{float}
\usepackage{stmaryrd}
\usepackage{graphicx}
\usepackage{caption}
\usepackage{subcaption}
\usepackage{authblk}
\usepackage[all]{xy}
\usepackage{tikz-cd}
\usepackage{hyperref}
\usepackage{cleveref}
\usepackage{multirow}
\usepackage{amsthm}
\usepackage{grffile}
\usepackage{comment}

\providecommand{\abs}[1]{\lvert#1\rvert}

\providecommand{\Sprod}[2]{\left\langle#1,#2\right\rangle}

\newenvironment{system}
  {\left\{ 
  \begin{aligned} 
  }
  {
  \end{aligned}
  \right.
  }

\newcommand{\uu}{\boldsymbol{u}}
\newcommand{\mm}{\boldsymbol{m}}
\newcommand{\vv}{\boldsymbol{v}}
\newcommand{\ww}{\boldsymbol{w}}

\newcommand{\Ab}{\boldsymbol{A}}
\newcommand{\Bb}{\boldsymbol{B}}
\newcommand{\Eb}{\boldsymbol{E}}
\newcommand{\Jb}{\boldsymbol{J}}

\newcommand{\bmu}{\boldsymbol{\mu}}

\newcommand{\bgamma}{\boldsymbol{\gamma}}
\newcommand{\blambda}{\boldsymbol{\lambda}}

\newcommand{\tlambda}{\tilde{\lambda}}

\newcommand{\tblambda}{\tilde{\blambda}}

\newcommand{\dxi}{{\partial_i}}
\newcommand{\dxj}{{\partial_j}}

\providecommand{\diff}[2]{\frac{\delta #1}{\delta #2}}

\DeclareMathOperator{\Div}{div}

\newtheorem{theorem}{Theorem}
\newtheorem{remark}{Remark}%

\title{A least Action principle for visco-resistive Hall Magnetohydrodynamic with metriplectic reformulation}
\author{Valentin Carlier, Martin Campos-Pinto}
\date{\today}

\begin{document}

\maketitle

\begin{abstract}
  We present a new variational formulation for Viscous and resistive Hall Magnetohydrodynamic. 
  We first find a variational principle for ideal HMHD by applying the physical assumptions leading to HMHD at the lagrangian level, and then we add the viscous and resistive terms by the means of constrained variations.
  We also provide a metriplectic reformulation of our formulation, based on two canonical Lie-Poisson brackets for the ideal part and metric 4-brackets for the dissipative part.
\end{abstract}

\section{Introduction}
Since the end of the 18th century, mathematicians and physicists have been describing numerous 
physical systems by means of variational principles.
In such a paradigm the central element is no longer the equation of motion per se, but rather the whole curve taken by the dynamical variables in phase space, characterized as the extremum of a certain functional which may be a Lagrangian or action functional. We refer to \cite{marsden2013introduction} for a general introduction.
This approach has proved extremely fruitful and has been used to build modern physics theories such as general relativity, quantum or particle physics \cite{landau2013quantum,berestetskii1982quantum}.
In particular, variational formulations are very appealing as they allow for an easier study of invariants of a given model (which appear as symmetries of the Lagrangian functional) or of its equilibria (which are extrema of the Lagrangian).
As such they have provided an efficient framework to derive stable 
numerical approximations \cite{courant_variational_1943, pavlov2011structure}: 
schemes based on discrete variational principles indeed benefit from very good stability properties, as they preserve by construction many discrete invariants of the system, 
see, e.g., \cite{gawlik2011geometric,kraus2017gempic,gawlik2021variational,carlier2025variational}.

After the first variational principles were formulated for incompressible and compressible fluids \cite{lagrange_mechanique_1788, serrin_mathematical_1959, arnold_sur_1966}, the approach was applied to the Vlasov-Maxwell equations in \cite{low_lagrangian_1958} and to ideal magnetohydrodynamics (MHD) in \cite{newcomb1961lagrangian}. 
Since then, several Lagrangian functionals have been proposed for extended 
and reduced MHD models such as Hall and electron MHD \cite{holm_hall_1987,ilgisonis_lagrangian_1999,yoshida_canonical_2013,keramidas_charidakos_action_2014}.

Let us note that many of these variational principles are formulated in Lagrangian coordinates, in the sense that they involve the characteristic flow as 
a dynamical variable. Although this corresponds to the historical form of the 
least action principle \cite{lagrange_mechanique_1788}
and is very suitable to derive variational approximations of particle type
\cite{campos_pinto_variational_2022}, Eulerian variables (associated with a fixed frame) 
are generally preferred in the numerical modelling of fluids. 
Because least action principles then involve Eulerian velocity fields as dynamical 
variables, they must be recast in a constrained form as was already observed 
in \cite{newcomb1961lagrangian}:
each fluid velocity must be optimized under constrained variations and the 
fluid densities become advected parameters. 
We refer to \cite{holm1998euler} for a detailed presentation of the corresponding Euler-Poincaré reduction theory.

In this work, our primary objective was to propose a variational principle for 
a visco-resistive MHD model. Traditionally indeed, least action principles are only able to describe non-dissipative systems, namely systems where all the forces derive from potentials, hence they cannot encompass systems with friction or heat transfer. 
Efforts have been made to add those phenomena to the scope of variational principles: among them we may cite the principle of least dissipation of energy \cite{onsager1931reciprocal} or of minimum entropy production \cite{prigogine1947etude,gyarmati1970non}. 
Here we have decided to follow the recent developments of 
\cite{gay2017lagrangian1,gay2017lagrangian2} where a generalized 
Lagrange-d'Alembert principle is proposed for non-equilibrium thermodynamics.
By using entropy production to enforce energy preservation, this new 
approach allows to handle any kind of dissipative force.
Motivated by the design of Eulerian numerical methods, we further follow 
\cite{holm1998euler, gawlik2021variational} and 
consider constrainted variational principles in Eulerian variables.

Another motivation was to derive an equivalent formulation in the framework of metriplectic systems 
\cite{kaufman1984dissipative,morrison1984bracket,morrison1986paradigm}
which aims at adding dissipation to initially conservative Hamiltonian systems.
It is well known that solutions to the equations given by extremizing a Lagrangian correspond to curves generated by a Hamiltonian and a Poisson bracket.
In this new paradigm, one first identifies a Casimir (a special invariant of the system, usually the entropy) that will be dissipated by the new dynamic,
then a second bracket is introduced
(either a 2-bracket in the earliest formulations, 
or a 4-bracket in more recent works \cite{morrison2024inclusive}):
This ``metric'' bracket preserves the Hamiltonian but dissipates the entropy.
Like the variational principle, this framework has been used with great success to derive numerical schemes that preserve physical invariants \cite{sanz1992symplectic,kraus2017metriplectic}.

In this article, we propose a first variational principle for
visco-resistive Hall-MHD equations, expressed in Eulerian variables. 
Our Lagrangian functional is derived in two steps: 
we first obtain a variational principle 
for the ideal case by expressing the physical assumptions 
of the Hall regime in a two-fluid Lagrangian with electromagnetic coupling,
and we next use the dissipative framework of \cite{gay2017lagrangian1,gay2017lagrangian2}
to incorporate viscosity and resistivity effects.
In particular, the latter is introduced via a force mimicking collisions 
between ions and electrons reminiscent of the Drude model, 
forcing the two flows to get closer to each other.
For this visco-resistive model we also provide an equivalent metriplectic rewriting, 
following the general derivation proposed in \cite{carlier2024metriplectic}.

The remainder of the article is organized as follows: in \cref{sec:two_fluid} 
we remind the variational principle for simple fluids in Eulerian variables, 
with an electromagnetic coupling. Then in \cref{sec:hmhd} we simplify the 
two-fluid model by assuming zero electron inertia and pressure, as well
as quasi-neutrality, leading to a variational principle for ideal Hall-MHD. 
In \cref{sec:vrhmhd} we present our variational viscoresistive 
model obtained by adding dissipative terms,
and specify its metriplectic reformulation.
\Cref{sec:conclusion} gathers some concluding remarks.

\section{Lagrangian variational principle for charged fluids in an electromagnetic field}
\label{sec:two_fluid}
We start our derivation by recalling the variational principle underlying compressible fluid dynamics
\cite{holm1998euler}, which we next couple with that of Maxwell's equations to obtain a model for a charged fluid. By combining the Lagrangian functionals associated with two species we then obtain a variational principle for two charged fluid interacting through an electromagnetic field: from this model we will be able to derive a Lagrangian for the Hall-MHD system.

\subsection{Euler equation from variational principle}

The variational principle for compressible fluid in standard (Eulerian) variables is obtained after reduction of a general variational principle set on the group if diffeomorphisms $D(\Omega)$, where $\Omega$ is the domain in which the fluid evolves. This results in a least action principle where variations in Eulerian variables are not free, but constrained by relations involving the Lie brackets of vector fields and general advection of density, known as Lin constraints. Here we only state and study the variational principles after reduction, and refer to \cite{holm1998euler} for a more detailed descrption of the Euler-Poincaré reduction. We denote $X(\Omega)$ the set of (smooth) vector fields of $\Omega$ that are tangent to the boundary, and $F(\Omega)$ the set of (smooth) functions on $\Omega$.

\begin{theorem}
\label{thm:var_prin_euler}
Consider the following Lagrangian  :
\begin{equation}
\label{eqn:lag_euler}
l(\uu, n, s) = \int_\Omega \frac{m_s n|\uu|^2}{2} - n e(n, s) ~,
\end{equation} 
where $\uu \in X(\Omega)$ , $n, s \in F(\Omega)$ and $e$ is the internal energy of the fluid, as a function of $n$ the density and $s$ the entropy. $m_s$ is the mass of a particle constituting the fluid.
Then extremizers of the action $\mathfrak{S} = \int_0^T l dt$ under constrained variations 
\begin{equation} 
  \label{eqn:euler_constraints}
    \delta \uu = \partial_t \vv + [\uu,\vv], 
    \qquad 
    \delta n = -\Div (n \vv),
    \qquad
    \delta s = -\Div (s \vv)
\end{equation}
(where 
$\vv = \vv(t)$ is a curve in $X(\Omega)$, i.e. a time dependent vector field, vanishing at $t=0$ and $T$, and $[\uu,\vv] = \uu \cdot \nabla \vv - \vv \cdot \nabla \uu$ is the Lie bracket of vector fields)
supplemented with the advection equations
\begin{equation}
\label{eqn:euler_advection}
\partial_t n + \Div(n \uu) = 0 ~,
\qquad
\partial_t s + \Div(s \uu) = 0 ~,
\end{equation}
are solution to the following system of equations 
\begin{subequations}
\label{eqn:euler}
\begin{equation}
\label{eqn:euler_mass}
\partial_t n + \Div(n \uu) = 0 ~,
\end{equation}
\begin{equation}
\label{eqn:euler_s}
\partial_t s + \Div(s \uu) = 0 ~,
\end{equation}
\begin{equation}
\label{eqn:euler_mom}
m_s n \partial_t \uu + m_s n (\uu \cdot \nabla \uu) + \nabla p = 0 ~, 
\end{equation}
\end{subequations}
where $p = p(n, s)$ is the associated pressure, given by $p=n(n \partial_{n} e+ s \partial_s e)$.
\end{theorem}
\begin{proof}
The first two equations are simply the advection equations. To find \cref{eqn:euler_mom}, we write the variational principle :
\begin{align*}
0 = \delta \mathfrak{S} &= \int_0^T \int_\Omega \delta l \\
	&= \int_0^T \int_\Omega \frac{\delta l}{\delta u}\delta u + \frac{\delta l}{\delta n}\delta n + \frac{\delta l}{\delta s}\delta s \\
	&= \int_0^T \int_\Omega \frac{\delta l}{\delta u}(\partial_t \vv + [\uu,\vv]) - \frac{\delta l}{\delta n}\Div(n \vv) - \frac{\delta l}{\delta s}\Div(s \vv) \\
	&= \int_0^T \int_\Omega m_s n \uu \cdot (\partial_t \vv + [\uu,\vv]) - (m_s \frac{|\uu|^2}{2} - \partial_n (n e)) \Div(n \vv) + \partial_s (n e)\Div(s \vv). 
\end{align*}
Using integration by parts we develop the first term as
\begin{align*}
        &\int_0^T \int_\Omega m_s n \uu \cdot (\partial_t \vv + \uu \cdot \nabla \vv - \vv \cdot \nabla \uu)
        \\
        &= - m_s \int_0^T \int_\Omega \partial_t(n \uu) \cdot \vv + n (\uu \cdot \nabla \uu) \cdot \vv + \uu \Div(n \uu) \cdot \vv  + (\nabla \uu)^T (n \uu) \cdot \vv,
\end{align*}
the second term as
\begin{align*}
        & - \int_0^T \int_\Omega \Big( m_s \frac{|\uu|^2}{2}-e(n, s)- n \partial_{n} e(n, s)\Big) \Div(n \vv)
        \\
        &=  \int_0^T \int_\Omega \nabla \Big( m_s \frac{|\uu|^2}{2} -e(n, s)-n \partial_{n} e(n, s)\Big) \cdot (n \vv)
        \\
        &=  \int_0^T \int_\Omega \big( m_s (\nabla \uu)^T (\uu) - \nabla n \partial_{n} e(n, s) - \nabla s \partial_{s} e(n, s) - \nabla (n \partial_{n} e(n, s)) \big) \cdot (n \vv)
\end{align*}
and the third term as
\begin{align*}
        \int_0^T \int_\Omega n \partial_s e(n, s) \Div (s \vv)
        =  - \int_0^T \int_\Omega \nabla (n \partial_s e(n, s)) \cdot (s \vv).
\end{align*}
Using $p=n(n \partial_{n} e+ s \partial_s e)$ and 
\cref{eqn:euler_advection} we next observe that
\begin{equation}
\label{eqn:id_pres}
n \big(\nabla n \partial_{n} e(n, s) + \nabla s \partial_{s} e(n, s) + \nabla (n \partial_{n} e(n, s))\big) + s \nabla (n \partial_s e(n, s))
     = \nabla p ~, 
\end{equation}
\begin{equation}
\label{eqn:id_u}
\partial_t(n \uu) + \uu\Div(n \uu) = \uu \partial_t n + n \partial_t (\uu) + \uu\Div(n \uu) = n \partial_t \uu  ~, 
\end{equation}
so that summing the three contributions above yields
\begin{equation*} 
0 = \int_0^T \int_\Omega \Big( - m_s n (\partial_t \uu - \uu \cdot \nabla \uu) - \nabla p \Big) \cdot \vv.
\end{equation*}
Equation \eqref{eqn:euler_mom} follows from the fact that the latter holds for all $\vv$ vanishing on the boundary.
\end{proof}

\subsection{Single charged fluid in an electromagnetic field}

A variational principle for a charged fluid can next be obtained by coupling the above fluid least action principle with an electromagnetic Lagrangian. The latter involves a potential $\phi$ and a vector potential $\Ab$, and corresponds to the energy of the electric field minus the one of the magnetic field. The coupling term is standard for particles in an electromagnetic field, considering here the fluid as a particle distribution. We also denote $q_s$ the electric charge of an individual particle of species $s$, so that $q_s n$ is the charge density.
\begin{theorem}
\label{thm:single_fluid}
Consider the following lagrangian :
\begin{subequations}
\label{eqn:lagrangians_single_fluid}
\begin{equation}
\label{eqn:lagrangian_single_fluid}
l(\uu, n, s, \Ab, \phi) = l_f(\uu, n, s, \Ab, \phi) + l_{max}(\Ab, \phi)
\end{equation}
\begin{equation}
\label{eqn:lagrangian_fluid}
l_f(\uu, n, s, \Ab, \phi) = \int_\Omega \frac{m_s n |\uu|^2}{2} - n e(n, s) + q_s n \uu \cdot \Ab - q_s n \phi 
\end{equation}
\begin{equation}
\label{eqn:lagrangian_full_maxwell}
l_{max}(\Ab, \phi) = \int_\Omega \epsilon_0 \frac{|\nabla \phi + \partial_t \Ab |^2}{2}- \frac{|\nabla \times \Ab|^2}{2 \mu_0}.
\end{equation}
\end{subequations}
Then extremizers of the action $\mathfrak{S} = \int_0^T l dt$ under constrained variations 
\begin{equation} 
  \label{constraints}
    \delta \uu = \partial_t \vv + [\uu,\vv], 
    \qquad 
    \delta n = -\Div (n \vv),
    \qquad
    \delta s = -\Div (s \vv),
\end{equation}
supplemented with the advection equations
\begin{equation}
\label{eqn:advection_single_fluid}
\partial_t n + \Div(n \uu) = 0 ~,
\qquad
\partial_t s + \Div(s \uu) = 0 ~,
\end{equation}
are solution to the following system of equations : 
\begin{subequations}
\label{eqn:single_fluid}
\begin{equation}
\label{eqn:sf_mass}
\partial_t n + \Div(n \uu) = 0 ~,
\end{equation}
\begin{equation}
\label{eqn:sf_s}
\partial_t s + \Div(s \uu) = 0 ~,
\end{equation}
\begin{equation}
\label{eqn:sf_mom}
m_s n \partial_t \uu + m_s n (\uu \cdot \nabla \uu) + \nabla p - q_s n (\Eb + \uu \times \Bb)  = 0 ~, 
\end{equation}
\begin{equation}
\label{eqn:sf_Ampere}
\epsilon_0 \partial_t \Eb = \frac{\nabla \times \Bb}{\mu_0} - q_s n \uu ~,
\end{equation}
\begin{equation}
\label{eqn:sf_Gauss}
\epsilon_0 \nabla \cdot \Eb = q_s n~,
\end{equation}
\end{subequations}
where $p$ is defined as in \cref{thm:var_prin_euler}, $E = -\nabla \phi -\partial_t \Ab$ and $\Bb = \nabla \times \Ab$.
\end{theorem}
\begin{proof}
The two first equations follow directly from the advection equations we consider, for the last three, we write the variational principle : $0 = \delta \mathfrak{S} $ with
\begin{align*}
\delta \mathfrak{S} &= \int_0^T \delta l \\
	&= \int_0^T \delta l_f + \delta l_{max} \\
	&= \int_0^T \frac{\delta l_f}{\delta u}\delta u + \frac{\delta l_f}{\delta n}\delta n + \frac{\delta l_f}{\delta s}\delta s + \frac{\delta l_f}{\delta \phi}\delta \phi + \frac{\delta l_f}{\delta \Ab}\delta \Ab + \frac{\delta l_{max}}{\delta \phi}\delta \phi + \frac{\delta l_{max}}{\delta \Ab}\delta \Ab \\
	&= \int_0^T \frac{\delta l_f}{\delta u}(\partial_t \vv + [\uu,\vv]) - \frac{\delta l_f}{\delta n}\Div(n \vv) - \frac{\delta l_f}{\delta s}\Div(s \vv) + \frac{\delta l_f}{\delta \phi}\delta \phi + \frac{\delta l_f}{\delta \Ab}\delta \Ab + \frac{\delta l_{max}}{\delta \phi}\delta \phi + \frac{\delta l_{max}}{\delta \Ab}\delta \Ab.
\end{align*}
Since variation in $\Ab$ and $\phi$ are independent of $\vv$, we may assume $\delta \Ab =0$ and $\delta \phi = 0$. In particular we may consider only variations implying $\vv$. 
Hence $\delta \mathfrak{S} = 0$ for all variations implies : 
\begin{equation}
\label{eqn:derivation_deltav}
\begin{aligned}
0 &= \int_0^T \frac{\delta l_f}{\delta u}(\partial_t \vv + [\uu,\vv]) - \frac{\delta l_f}{\delta n}\Div(n \vv) - \frac{\delta l_f}{\delta s}\Div(s \vv) \\
	&= 
  (I) + (II) + (III)
\end{aligned}
\end{equation}
where the first term is
(using integration by parts as in the proof of \cref{thm:var_prin_euler})
\begin{align*}
  (I) &= \int_0^T (m_s n \uu + q_s n A)(\partial_t \vv + [\uu,\vv])
    \\
      &=
        -\int_0^T \int_\Omega \partial_t(m_s n \uu + q_s n \Ab) \cdot \vv + n (\uu \cdot \nabla (m_s \uu + q_s \Ab)) \cdot \vv + (m_s \uu + q_s \Ab)\Div( n \uu) \cdot \vv  
        \\
        &\qquad \qquad  
        + (\nabla \uu)^T (n (m_s \uu+ q_s \Ab)) \cdot \vv
\end{align*}
the second one is
\begin{align*}
  (II) &= - \int_0^T  (m_s \frac{|u|^2}{2} - \partial_n (n e) 
            - q_s \phi + q_s \uu \cdot \Ab) \Div(n \vv) 
  \\
  &=
        \int_0^T \int_\Omega \big( m_s (\nabla \uu)^T (\uu) - \nabla n \partial_{n} e(n, s) - \nabla s \partial_{s} e(n, s) - \nabla (n \partial_{n} e(n, s)) + q_s \nabla (\Ab \cdot \uu) - q_s \nabla \phi \big) \cdot ( \vv)
\end{align*}
and the third is
\begin{align*}
    (III) &= \int_0^T  \partial_s (n e)\Div(s \vv)
    =   - \int_0^T \int_\Omega \nabla (n \partial_s e(n, s)) \cdot (s \vv).
\end{align*}
We next observe that
\begin{equation}
\label{eqn:id_rho}
\nabla(\Ab \cdot \uu) - \uu \cdot \nabla \Ab - \nabla \uu ^T \Ab = \uu \times \nabla \times \Ab ~, 
\end{equation}
so that summing the above three contributions and using the latter identity, \cref{eqn:id_pres}, and a version of \cref{eqn:id_u} where $\uu$ is replaced by $m_s \uu + q_s \Ab$, we find that
\begin{equation} \label{eqn:sf_mom_v}
0 = \int_0^T \int_\Omega \Big( - n \partial_t (m_s \uu + q_s \Ab) - m_s n \uu \cdot \nabla \uu - \nabla p - q_s n \nabla \phi + m_s n \uu \times \nabla \times \Ab \Big) \cdot \vv .
\end{equation}
Again the claimed momentum equation \cref{eqn:sf_mom} follows 
from the fact that \eqref{eqn:sf_mom_v} holds for all $\vv$ null at the boundaries, 
und by using the definitions for $\Eb$ and $\Bb$.
%
%

Taking next variations in $\delta \Ab$, we write 
\begin{align*}
0 &= \int_0^T \int_\Omega \frac{\delta l_f}{\delta \Ab}\delta \Ab + \frac{\delta l_{max}}{\delta \Ab}\delta \Ab \\
  &= \int_0^T \int_\Omega q_s n \uu \cdot \delta \Ab + \epsilon_0 (\nabla \phi + \partial_t \Ab) \cdot \partial_t \delta \Ab - \frac{1}{\mu_0}\nabla \times \Ab \cdot \nabla \times \delta \Ab  \\
  &= \int_0^T \int_\Omega \big( q_s n \uu - \epsilon_0 \partial_t (\nabla \phi + \partial_t \Ab) - \frac{1}{\mu_0}\nabla \times \nabla \times \Ab \big) \cdot \delta \Ab ~.
\end{align*}
So that we obtain \cref{eqn:sf_Ampere} since the above equality holds for every $\delta \Ab$ that is null at the boundaries.

Finally for the variations in $\delta \phi$, we write
\begin{align*}
0 &= \int_0^T \int_\Omega \frac{\delta l_f}{\delta \phi}\delta \phi + \frac{\delta l_{max}}{\delta \phi}\delta \phi \\
  &= \int_0^T \int_\Omega -q_s n \delta \phi + \epsilon_0 (\nabla \phi + \partial_t \Ab) \cdot \nabla \delta \phi \\
  &= \int_0^T \int_\Omega -q_s n \delta \phi - \epsilon_0 \nabla \cdot (\nabla \phi + \partial_t \Ab) \delta \phi ~,
\end{align*}
which yields the equation \cref{eqn:sf_Gauss}. 
\end{proof}

\subsection{Two charged fluids in an electromagnetic field}
We finish this section by presenting a variational formulation for a mixture of two charged fluids. In this first model the two fluids only interact through the electromagnetic field: it does not bring any conceptual novelty compared to a single charged fluid (in fact we only sum the two fluid Lagrangians without changing the Maxwell part), it is nevertheless
useful as a basis for the Hall magnetohydrodynamics model that will be presented next.
Without loss of generality we denote the two species with subscripts $e$ and $i$, for electrons and ions.
\begin{theorem}
\label{thm:two_fluid}
Consider the following lagrangian :
\begin{subequations}
\label{eqn:lagrangians_tfluid}
\begin{equation}
\label{eqn:lagrangian_tfluid}
l(\uu_i, n_i, s_i, u_e, n_e, s_e, \Ab, \phi) = l_i(\uu_i, n_i, s_i, \Ab, \phi) + l_e(\uu_e, n_e, s_e, \Ab, \phi) + l_{max}(\Ab, \phi)
\end{equation}
\begin{equation}
\label{eqn:lagrangian_ion_tfluid}
l_i(\uu_i, n_i, s_i, \Ab, \phi) = \int_\Omega \frac{m_i n_i|\uu_i|^2}{2} - n_i e_i(n_i, s_i) + q_i n_i \uu_i \cdot \Ab - q_i n_i \phi
\end{equation}
\begin{equation}
\label{eqn:lagrangian_elec_tfluid}
l_e(\uu_e, n_e, s_e, \Ab, \phi) = \int_\Omega \frac{m_e n_e|\uu_e|^2}{2} - n_e e_e(n_e, s_e) + q_e n_e \uu_e \cdot \Ab - q_e n_e \phi
\end{equation}
\begin{equation}
\label{eqn:lagrangian_tfluid_maxwell}
l_{max}(\Ab, \phi) = \int_\Omega \epsilon_0 \frac{|\nabla \phi + \partial_t \Ab |^2}{2}- \frac{|\nabla \times \Ab|^2}{2 \mu_0}
\end{equation}
\end{subequations}
Then extremizers of the action $\mathfrak{S} = \int_0^T l dt$ under constrained variations 
\begin{align} 
  \label{eqn:tfluid_constraints}
    \delta \uu_i = \partial_t \vv_i + [\uu_i,\vv_i], 
    \qquad 
    \delta n_i = -\Div (n_i \vv_i),
    \qquad
    \delta s_i = -\Div (s_i \vv_i) \\
    \delta \uu_e = \partial_t \vv_e + [\uu_e,\vv_e], 
    \qquad 
    \delta n_e = -\Div (n_e \vv_e),
    \qquad
    \delta s_e = -\Div (s_e \vv_e)
\end{align}
supplemented with the advection equations :
\begin{subequations}
\begin{equation}
\label{eqn:advection_rhoi}
\partial_t n_i + \Div(n_i \uu_i) = 0 ~,
\end{equation}
\begin{equation}
\label{eqn:advection_si}
\partial_t s_i + \Div(s_i \uu_i) = 0 ~.
\end{equation}
\begin{equation}
\label{eqn:advection_rhoe}
\partial_t n_e + \Div(n_e \uu_e) = 0 ~,
\end{equation}
\begin{equation}
\label{eqn:advection_se}
\partial_t s_e + \Div(s_e \uu_e) = 0 ~.
\end{equation}
\end{subequations}
Are solution to the following system of equations 
\begin{subequations}
\label{eqn:two_fluids}
\begin{equation}
\label{eqn:tf_massi}
\partial_t n_i + \Div(n_i \uu_i) = 0 ~,
\end{equation}
\begin{equation}
\label{eqn:tf_si}
\partial_t s_i + \Div(s_i \uu_i) = 0 ~,
\end{equation}
\begin{equation}
\label{eqn:tf_masse}
\partial_t n_e + \Div(n_e \uu_e) = 0 ~,
\end{equation}
\begin{equation}
\label{eqn:tf_se}
\partial_t s_e + \Div(s_e \uu_e) = 0 ~,
\end{equation}
\begin{equation}
\label{eqn:tf_momi}
m_i n_i ( \partial_t \uu_i + (\uu_i \cdot \nabla \uu_i)) + \nabla p_i - q_i n_i (\Eb + \uu_i \times \Bb)  = 0 ~, 
\end{equation}
\begin{equation}
\label{eqn:tf_mome}
m_e n_e ( \partial_t \uu_e + (\uu_e \cdot \nabla \uu_e)) + \nabla p_e - q_e n_e (\Eb + \uu_e \times \Bb)  = 0 ~, 
\end{equation}
\begin{equation}
\label{eqn:tf_Ampere}
\epsilon_0 \partial_t \Eb = \frac{\nabla \times \Bb}{\mu_0} - q_i n_i \uu_i - q_e n_e \uu_e ~,
\end{equation}
\begin{equation}
\label{eqn:tf_Gauss}
\epsilon_0 \nabla \cdot \Eb = q_i n_i  + q_e n_e ~,
\end{equation}
\end{subequations}
\end{theorem}
\begin{proof}
The proof is similar to the one of \cref{thm:single_fluid}: the advection equations are imposed, the equation for $\uu_i$ (resp. $\uu_e$) follows by taking variations in 
$\delta n_i$, $\delta s_i$ and $\delta \uu_i$ (resp. $\delta n_e$, $\delta s_e$ and $\delta \uu_e$) and summing all contributions involving $\vv_i$ (resp. $\vv_e$). Finally the equations for $\Eb$ and $\Bb$ are obtained by taking variations in $\delta \Ab$ and $\delta \phi$.
\end{proof}

\section{Hall-MHD}
\label{sec:hmhd}
\subsection{From two charged fluids to Hall MHD}

The equations for Hall MHD can be derived from the two-fluid model above by making three assumptions: the electrons are massless ($m_e=0$), they are pressureless 
($e_e=0$ so $p_e=0$) 
and the fluid is quasineutral 
(which is often imposed by formally letting $\epsilon_0 = 0$). 
The first two assumptions allow to rewrite \cref{eqn:tf_mome} as 
\begin{equation}
\label{eqn:hmhd_mome_intro}
\Eb + \uu_e \times \Bb = 0 ~, 
\end{equation}
while the last assumption transforms \cref{eqn:tf_Ampere,eqn:tf_Gauss} in 
\begin{equation*}
\frac{\nabla \times \Bb}{\mu_0} = q_i n_i \uu_i + q_e n_e \uu_e =: \Jb
\qquad \text{ and } \qquad
0 = q_i n_i  + q_e n_e ~,
\end{equation*}
which can be combined into 
\begin{equation*}
\uu_e = \frac{1}{q_e n_e} \Jb + \uu_i
~.
\end{equation*}
Injecting this relation in \cref{eqn:hmhd_mome_intro} we find the Ohm law of Hall MHD : 
\begin{equation}
\label{eqn:Hall_Ohm_law}
\Eb = 
\Big(\frac{1}{q_i n_i} \Jb - \uu_i\Big) \times \Bb ~,
\end{equation}
and substituting in \cref{eqn:tf_momi} we recover the standard momentum equation
\begin{equation}
\label{eqn:Hall_mom}
m_i n_i (\partial_t \uu_i + (\uu_i \cdot \nabla \uu_i)) 
= \Jb \times \Bb - \nabla p_i ~.
\end{equation}
Moreover, Faraday's equation follows from the definition of $\Eb$ and $\Bb$:
\begin{equation}
\label{eqn:Hall_Faraday}
\partial_t \Bb = \partial_t (\nabla \times \Ab) = \nabla \times (\partial_t \Ab) = \nabla \times (\partial_t \Ab + \nabla \phi) = - \nabla \times \Eb 
= \nabla \times \left( \Big(-\frac{\Jb}{q_i n_i}+ \uu_i\Big)  \times \Bb \right) ~.
\end{equation}
Finally the remaining equations (advections of $n$ and $s$) are left unchanged, so that we indeed recover the set of Hall MHD equations, see e.g. \cite{d2016derivation}.

\subsection{A first variational principle for Hall-MHD}

By transcribing our three hypotheses in the two-fluid Lagrangian 
we obtain a first variational principle for the equations of Hall MHD. 
To do so we directly set $m_e=0$, $e_e = 0$ and $\epsilon_0=0$ in the Lagrangian from \cref{thm:two_fluid} and derive the new equations satisfied by the extremizers. Note that the electron Lagrangian no longer depends on $s_e$, 
hence we can remove this term from the dynamical variables.
\begin{theorem}
\label{thm:hmhd}
Consider the following lagrangian :
\begin{subequations}
\label{eqn:lagrangians_hmhd}
\begin{equation}
\label{eqn:lagrangian_hmhd}
l(\uu_i, n_i, s_i, \uu_e, n_e, \Ab, \phi) = l_i(\uu_i, n_i, s_i, \Ab, \phi) + l_e(\uu_e, n_e, \Ab, \phi) + l_{max}(\Ab)
\end{equation}
\begin{equation}
\label{eqn:lagrangian_ion_hmhd}
l_i(\uu_i, n_i, s_i, \Ab, \phi) = \int_\Omega \frac{m_i n_i|\uu_i|^2}{2} - n_i e_i(n_i, s_i) + q_i n_i \uu_i \cdot \Ab - q_i n_i \phi 
\end{equation}
\begin{equation}
\label{eqn:lagrangian_elec_hmhd}
l_e(\uu_e, n_e, \Ab, \phi) = \int_\Omega q_e n_e \uu_e \cdot \Ab - q_e n_e \phi
\end{equation}
\begin{equation}
\label{eqn:lagrangian_hmhd_maxwell}
l_{max}(\Ab) = \int_\Omega - \frac{|\nabla \times \Ab|^2}{2 \mu 0}~.
\end{equation}
\end{subequations}
Then extremizers of the action $\mathfrak{S} = \int_0^T l dt$ under constrained variations 
\begin{align} 
  \label{eqn:hmhd_constraints}
    &\delta \uu_i = \partial_t \vv_i + [\uu_i,\vv_i], 
    \qquad 
    \delta n_i = -\Div (n_i \vv_i),
    \qquad
    \delta s_i = -\Div (s_i \vv_i) \\
    &\delta \uu_e = \partial_t \vv_e + [\uu_e,\vv_e], 
    \qquad 
    \delta n_e = -\Div (n_e \vv_e),
\end{align}
supplemented with the advection equations :
\begin{subequations}
\begin{equation}
\label{eqn:advection_rhoi_hmhd}
\partial_t n_i + \Div(n_i \uu_i) = 0 ~,
\end{equation}
\begin{equation}
\label{eqn:advection_si_hmhd}
\partial_t s_i + \Div(s_i \uu_i) = 0 ~,
\end{equation}
\begin{equation}
\label{eqn:advection_rhoe_hmhd}
\partial_t n_e + \Div(n_e \uu_e) = 0 ~,
\end{equation}
\end{subequations}
are solution to the following system of equations 
\begin{subequations}
\label{eqn:hmhd}
\begin{equation}
\label{eqn:hmhd_massi}
\partial_t n_i + \Div(n_i \uu_i) = 0 ~,
\end{equation}
\begin{equation}
\label{eqn:hmhd_si}
\partial_t s_i + \Div(s_i \uu_i) = 0 ~,
\end{equation}
\begin{equation}
\label{eqn:hmhd_masse}
\partial_t n_e + \Div(n_e \uu_e) = 0 ~,
\end{equation}
\begin{equation}
\label{eqn:hmhd_momi}
m_i n_i (\partial_t \uu_i + \uu_i \cdot \nabla \uu_i) + \nabla p_i - q_i n_i (\Eb + \uu_i \times \Bb)  = 0 ~, 
\end{equation}
\begin{equation}
\label{eqn:hmhd_mome}
q_e n_e (\Eb + \uu_e \times \Bb)  = 0 ~, 
\end{equation}
\begin{equation}
\label{eqn:hmhd_Ampere}
0 = \frac{\nabla \times \Bb}{\mu_0} - q_i n_i \uu_i - q_e n_e \uu_e ~,
\end{equation}
\begin{equation}
\label{eqn:hmhd_Gauss}
0 = q_i n_i + q_e n_e ~.
\end{equation}
\end{subequations}
\end{theorem}
\begin{proof}
\Cref{eqn:hmhd_massi,eqn:hmhd_si,eqn:hmhd_masse,eqn:hmhd_momi} are obtained the same way as in the proof of \cref{thm:single_fluid}. 
For \cref{eqn:hmhd_mome} we consider variations in $\vv_e$, that is, in $\delta \uu_e$, and $\delta n_e$ :
\begin{align*}
0 &= \int_0^T \int_\Omega \frac{\delta l_e}{\delta \uu_e}\delta \uu_e + \frac{\delta l_e}{\delta n_e}\delta n_e \\
	&= \int_0^T \int_\Omega \frac{\delta l_e}{\delta \uu_e}(\partial_t \vv_e + [\uu_e,\vv_e]) - \frac{\delta l_e}{\delta n_e}\Div(n_e \vv_e)\\
	&= \int_0^T \int_\Omega (q_e n_e \Ab)(\partial_t \vv_e + [\uu_e,\vv_e]) - (q_e \phi + q_e \uu_e \cdot \Ab) \Div(n_e \vv_e) \\
	&= \int_0^T \int_\Omega -\partial_t (q_e n_e \Ab)\cdot \vv_e - q_e n_e (\uu_e \cdot \nabla \Ab) \cdot \vv_e -q_e \Div(n_e \uu_e) \Ab \cdot \vv_e - q_e n_e (\nabla \uu_e)^T \Ab \cdot \vv_e \\
	&+ q_e \nabla (\uu_e \cdot \Ab) \cdot (n_e \vv_e)  + q_e n_e (\nabla \phi) \cdot \vv_e \\
	&= \int_0^T \int_\Omega q_e n_e (-\partial_t \Ab - \nabla \phi + \uu_e \times \nabla \times \Ab) \cdot \vv_e ~, 
\end{align*}
using identities \cref{eqn:id_u,eqn:id_rho}, which is \cref{eqn:hmhd_mome}. Variations in $\delta \phi$ now give 
\begin{align*}
0 &= \int_0^T \int_\Omega \frac{\delta l_e}{\delta \phi}\delta \phi + \frac{\delta l_i}{\delta \phi}\delta \phi \\
  &= \int_0^T \int_\Omega - q_e n_e \delta \phi - q_i n_i \delta \phi ~,
\end{align*}
that is \cref{eqn:hmhd_Gauss}, and variations in $\delta A$ give :
\begin{align*}
0 &= \int_0^T \int_\Omega \frac{\delta l_e}{\delta \Ab}\delta \Ab + \frac{\delta l_i}{\delta \Ab}\delta \Ab + \frac{\delta l_{max}}{\delta \Ab}\delta \Ab \\
  &= \int_0^T \int_\Omega q_e n_e \uu_e \cdot \delta \Ab + q_i n_i \uu_i \cdot \delta \Ab - \frac{1}{\mu_0} \nabla \times \Ab \cdot \nabla \times \delta \Ab \\
  &= \int_0^T \int_\Omega q_e n_e \uu_e \cdot \delta \Ab + q_i n_i \uu_i \cdot \delta \Ab - \frac{1}{\mu_0} \nabla \times \nabla \times \Ab \cdot \delta \Ab ~,
\end{align*}
which yields \cref{eqn:hmhd_Ampere}.
\end{proof}

\subsection{Simplified Hall-MHD}

In this section we propose a simplified variational principle for the Hall-MHD equations, which 
involves only five dynamical variables. 
To do so we observe that in the previous Lagrangian formulation, $\phi$ plays the role of a Lagrange multiplier enforcing the neutrality equation \cref{eqn:hmhd_Gauss} and that the two densities $n_i$, $n_e$ are linked via this equation.
This allows us to simplify the Lagrangian by removing the variables $\phi$ and $n_e$
(and renaming $n = n_i$, $s =s_i$, $e = e_i$), while keeping the same equations of motion.

\begin{theorem}
\label{thm:shmhd}
Consider the following lagrangian :
\begin{equation}
\label{eqn:lagrangian_shmhd}
L(\uu_i, n, s, \uu_e, \Ab) = \int_\Omega \frac{m_i n |\uu_i|^2}{2} - n e(n, s) + q_i n (\uu_i-\uu_e) \cdot \Ab - \frac{|\nabla \times \Ab|^2}{2 \mu_0} ~.
\end{equation}
Then extremizers of the action $\mathfrak{S} = \int_0^T L dt$ under constrained variations 
\begin{align} 
  \label{eqn:shmhd_constraints}
    &\delta \uu_i = \partial_t \vv_i + [\uu_i,\vv_i], 
    \qquad 
    \delta n = -\Div (n \vv_i),
    \qquad
    \delta s = -\Div (s \vv_i) \\
    &\delta \uu_e = \partial_t \vv_e + [\uu_e,\vv_e], 
\end{align}
supplemented with the advection equations :
\begin{subequations}
\begin{equation}
\label{eqn:advection_rhoi_shmhd}
\partial_t n + \Div(n \uu_i) = 0 ~,
\end{equation}
\begin{equation}
\label{eqn:advection_si_shmhd}
\partial_t s + \Div(s \uu_i) = 0 ~,
\end{equation}
\end{subequations}
are solutions to the following system of equations 
\begin{subequations}
\label{eqn:shmhd}
\begin{equation}
\label{eqn:shmhd_mass}
\partial_t n + \Div(n \uu_i) = 0 ~,
\end{equation}
\begin{equation}
\label{eqn:shmhd_s}
\partial_t s + \Div(s \uu_i) = 0 ~,
\end{equation}
\begin{equation}
\label{eqn:shmhd_momi}
m_i n ( \partial_t \uu_i +  \uu_i \cdot \nabla \uu_i) + \nabla p - q_i n (\Eb + \uu_i \times \Bb)  = 0 ~, 
\end{equation}
\begin{equation}
\label{eqn:shmhd_mome}
q_i n (\Eb + \uu_e \times \Bb)  = 0 ~, 
\end{equation}
\begin{equation}
\label{eqn:shmhd_Ampere}
0 = \frac{\nabla \times \Bb}{\mu_0} - q_i n(\uu_i - \uu_e) ~,
\end{equation}
\end{subequations}
where $p = p_i$ is the ion pressure 
(defined as in theorem~\ref{thm:var_prin_euler})
and $\Eb = -\partial_t \Ab - \nabla \phi$ is defined by setting $\phi = \Ab \cdot \uu_e$.
\end{theorem}
\begin{remark}
  The Lagrangian \eqref{eqn:lagrangian_shmhd} 
  is formally the same as the one of Ilgisonis and Lakhin 
  \cite{ilgisonis_lagrangian_1999} 
  (see also \cite{ilgisonis_variational_2000}), but here 
  the dynamical variables are expressed in Eulerian coordinates.
\end{remark}
\begin{proof}
\Cref{eqn:shmhd_mass,eqn:shmhd_s} are again the given advection equations. Considering next the variations in $\delta \Ab$, we write
\begin{align*}
0 =& \int_0^T \int_\Omega q_i n (\uu_i - \uu_e) \cdot \delta \Ab - \frac{1}{\mu_0} \nabla \times \Ab \cdot \nabla \times \delta \Ab ~,
\end{align*}
which directly gives \cref{eqn:shmhd_Ampere}. Since now only $\delta \uu_e$ has variations in $\vv_e$, the corresponding term reads :
\begin{align*}
0 = \int_0^T \int_\Omega \frac{\delta L}{\delta \uu_e} \cdot \delta \uu_e
  &= \int_0^T \int_\Omega \frac{\delta L}{\delta \uu_e} \cdot (\partial_t \vv_e + [\uu_e, \vv_e]) \\
  &= \int_0^T \int_\Omega - q_i n \Ab \cdot (\partial_t \vv_e + [\uu_e, \vv_e])\\
  &= \int_0^T \int_\Omega (q_i \partial_t(n \Ab) + q_i n (\uu_e \cdot \nabla) \Ab + q_i \Ab \nabla \cdot (n \uu_e) + q_i n (\nabla \uu_e)^T \cdot \Ab) \cdot \vv_e ~.
\end{align*}
From \cref{eqn:shmhd_Ampere} we next infer that $\nabla \cdot (n \uu_i) = \nabla \cdot (n \uu_e)$, so that we can use the same identity as in \cref{eqn:id_rho,eqn:id_u}, and write
\begin{align*}
0 &= \int_0^T \int_\Omega (q_i n \partial_t \Ab + q_i n (\uu_e \cdot \nabla) \Ab + q_i n (\nabla \uu_e)^T \cdot \Ab) \cdot \vv_e \\
  &= \int_0^T \int_\Omega q_i n (\partial_t \Ab + \nabla (\Ab \cdot \uu_e) - \uu_e \times (\nabla \times \Ab)) \cdot \vv_e
  \\
  &= -\int_0^T \int_\Omega q_i n ( \Eb + \uu_e \times \Bb) \cdot \vv_e
\end{align*}
which is \cref{eqn:shmhd_mome}, by using the definition of 
$\Eb = -\partial_t \Ab - \nabla \phi$ with $\phi = \Ab \cdot \uu_e$
and $\Bb = \nabla \times \Ab$. 
Finally by taking variations in $\vv_i$ 
we recognize exactly the same equation as in \cref{eqn:derivation_deltav}: 
therefore the same computation gives \cref{eqn:shmhd_momi}.
\end{proof}

\subsection{Canonical Hamiltonian and bracket formulation}

Turning to the Hamiltonian point of view, we next derive a Hamiltonian formulation for the above  Hall-MHD equations. To do so we apply a standard Legendre transform, while the brackets are found by combining standard brackets for systems with advected parameters \cite{holm1998euler}.
\begin{theorem}
\label{thm:hmhd_symplectic}
Given the Lagrangian defined by \cref{eqn:lagrangian_shmhd}, 
let the associated canonical momenta
$$
\mm_i = \frac{\delta L}{\delta \uu_i}= n (m_i \uu_i + q_i \Ab) 
\quad \text{and} \quad 
\mm_e = \frac{\delta L }{\delta \uu_e} = -q_i n \Ab~.
$$
Then, a curve $t \to (\uu_i, n, s, \Ab, \uu_e)(t)$ is a solution to the variational equations in \cref{thm:shmhd} iff the corresponding curve $t \to (\mm_i, \mm_e, n, s)(t)$ solves the 
Hamiltonian system 
$$
\dot{F} = \{F, H\}
$$ 
with Hamiltonian functional 
$H = \langle \mm_i, \uu_i \rangle + \langle \mm_e, \uu_e \rangle - L(\uu_i, n, s, \uu_e, \Ab)$, namely
\begin{equation}
\label{eqn:hamiltonian_hmhd}
H(\mm_i, \mm_e, n, s) = 
\int_\Omega \frac{|\mm_i+\mm_e|^2}{2 m_i n} + n e(n, s) + \frac{1}{2  \mu_0 q_i^2}|\nabla \times \frac{\mm_e}{n}|^2 ~,
\end{equation} 
and bracket 
\begin{subequations}
  \label{eqn:brackets_hmhd}
\begin{equation}
\label{eqn:bracket_hmhd}
\{F,G\} = \{F,G\}_{\mm_i} + \{F,G\}_{\mm_e} + \{F,G\}_{n} + \{F,G\}_{s}
\end{equation}
defined by 
\begin{equation}
\label{eqn:bracket_hmhd_mi}
\{F,G\}_{\mm_i} = \int_\Omega \mm_i \cdot [\frac{\delta G}{\delta \mm_i}, \frac{\delta F}{\delta \mm_i}] ~,
\end{equation}
\begin{equation}
\label{eqn:bracket_hmhd_me}
\{F,G\}_{\mm_e} = \int_\Omega \mm_e \cdot [\frac{\delta G}{\delta \mm_e}, \frac{\delta F}{\delta \mm_e}] ~,
\end{equation}
\begin{equation}
\label{eqn:bracket_hmhd_rho}
\{F,G\}_{n} = \int_\Omega \frac{\delta G}{\delta n} \nabla \cdot (n \frac{\delta F}{\delta \mm_i}) - \frac{\delta F}{\delta n} \nabla \cdot (n \frac{\delta G}{\delta \mm_i}) ~,
\end{equation}
\begin{equation}
\label{eqn:bracket_hmhd_s}
\{F,G\}_{s} = \int_\Omega \frac{\delta G}{\delta s} \nabla \cdot (s \frac{\delta F}{\delta \mm_i}) - \frac{\delta F}{\delta s} \nabla \cdot (s \frac{\delta G}{\delta \mm_i}) ~.
\end{equation}
\end{subequations}
\end{theorem}
\begin{proof}
Solutions defined by \cref{thm:shmhd} satisfy 
\begin{equation*}
\int_\Omega \frac{d}{dt}\frac{\delta L}{\delta \uu_i} \cdot \vv_i = \int_\Omega \frac{\delta L}{\delta \uu_i} \cdot [\uu_i, \vv_i] - \frac{\delta L}{\delta n} \nabla \cdot (n \vv_i) - \frac{\delta L}{\delta s} \nabla \cdot (s \vv_i)  ~.
\end{equation*}
Now, since  $\frac{\delta L}{\delta \uu_i} = \mm_i$, $\frac{\delta L}{\delta n} = -\frac{\delta H}{\delta n}$, $\frac{\delta L}{\delta s} = -\frac{\delta H}{\delta s}$ and $\uu_i = \frac{\delta H}{\delta \mm_i}$, we can rewrite this as 
\begin{equation*}
\int_\Omega \dot{\mm_i} \cdot \vv_i = \int_\Omega \mm_i \cdot [\frac{\delta H}{\delta \mm_i}, \vv_i] + \frac{\delta H}{\delta n} \nabla \cdot (n \vv_i) + \frac{\delta H}{\delta s} \nabla \cdot (s \vv_i) ~.
\end{equation*}
We also have that $\uu_e= \frac{\delta H}{\delta \mm_e}$, so that the equation for variations in $\vv_e$ can be rewritten as 
\begin{equation*}
\int_\Omega \dot{\mm_e} \cdot \vv_e = \int_\Omega \mm_e \cdot [\frac{\delta H}{\delta \mm_e}, \vv_e]~.
\end{equation*}
Let us next consider a functional $F = F(\mm_i, \mm_e, n, s)$: we have 
\begin{align*}
\dot{F} &= \int_\Omega \frac{\delta F}{\delta \mm_i} \dot{\mm_i} + \frac{\delta F}{\delta \mm_e} \dot{\mm_e} + \frac{\delta F}{\delta n} \dot{n} + \frac{\delta F}{\delta s} \dot{s} \\
	    &= \int_\Omega \mm_i \cdot [\frac{\delta H}{\delta \mm_i}, \frac{\delta F}{\delta \mm_i}] + \frac{\delta H}{\delta n} \nabla \cdot (n \frac{\delta F}{\delta \mm_i}) + \frac{\delta H}{\delta s} \nabla \cdot (s \frac{\delta F}{\delta \mm_i}) \\
	    &+ \int_\Omega \mm_e \cdot [\frac{\delta H}{\delta \mm_e}, \frac{\delta F}{\delta \mm_e}] - \frac{\delta F}{\delta n} \nabla \cdot (n \frac{\delta H}{\delta \mm_i}) - \frac{\delta F}{\delta s} \nabla \cdot (s \frac{\delta H}{\delta \mm_i}) \\
	    &= \{F, H\}.
\end{align*}
The remaining equations are obtained by taking functionals of the form 
$F(\mm_i) = \int_\Omega \mm_i \cdot \vv_i$, $F(n) = \int_\Omega n v$ and so forth, 
for given test functions $\vv_i$ and $v$.
\end{proof}

Although the above Hamiltonian formulation is derived from a variational principle, 
it coincides with the standard ones \cite{d2016derivation,coquinot2020general} when expressed in variables
$(\uu, n, s, \Bb)$.

\subsection{Comparison with other Hamiltonian formulations}

Several Hamiltonian formulations have been proposed for the 
Hall-MHD equations. The one derived by Holm in \cite{holm_hall_1987} 
involves a bracket with $(\mm_i, n, s, n_e, \Ab)$ as dynamical variables.
In \cite{yoshida_canonical_2013}, Yoshida and Hameiri proposed a bracket expressed in $(\uu, \Bb, \rho)$ variables with
\begin{equation} \label{eqn:uB_vars}
  \begin{system}
    &\uu = (\mm_i+\mm_e)/(m_i n)~,
    \\
    &\Bb = \nabla \times \Ab 
      = - \nabla \times \Big(\frac {\mm_e}{q_i n}\Big)~,
    \\
    &\rho = m_i n~,
  \end{system} 
\end{equation}
and 
in \cite{coquinot2020general} Coquinot and Morrison 
derived a Hamiltonian formulation
in $(\mm, \Bb, n)$ variables, 
\begin{equation} \label{eqn:mB_vars}
  \begin{system}
    &\mm = \mm_i + \mm_e
    \\
    &\Bb = \nabla \times \Ab 
      = - \nabla \times \Big(\frac {\mm_e}{q_i n}\Big).
  \end{system} 
\end{equation}
The formulations in \cite{yoshida_canonical_2013} and \cite{coquinot2020general} have the same Hamiltonian functional,
which coincide with ours. 
Moreover, denoting
$$
\tilde{f}(\tilde{y}) := f(y) := F(Y)
\qquad \text{ and } \qquad 
\tilde{g}(\tilde{y}) := g(y) := G(Y)
$$
the functions corresponding to the change of variables 
$$
Y = (\mm_i, \mm_e, n, s) \to y = (\mm, \Bb, n, s)
\to \tilde{y} = (\uu, \Bb, \rho, s),
$$
we compute
\begin{equation} \label{eqn:chgvar}
\begin{system}
  &\diff{F}{\mm_i} = \diff{f}{\mm} 
  \\
  &\diff{F}{\mm_e} = \diff{f}{\mm} - \frac {1}{q_i n} \nabla \times \diff{f}{\Bb}
  \\
  &\diff{F}{n} = \diff{f}{n} -\frac {\Ab }{n}\cdot \nabla \times \diff{f}{\Bb}
  \\
  &\diff{F}{s} = \diff{f}{s}
\end{system} 
\qquad \text{ and } \qquad 
\begin{system}
  &\diff{f}{\mm} = \frac{1}{\rho}\diff{\tilde{f}}{\uu},
  \\
  &\diff{f}{n} = m_i (\diff{\tilde{f}}{\rho} - \frac{\uu}{\rho}\diff{\tilde{f}}{\uu})~.
\end{system} 
\end{equation}
In particular, our bracket rewrites as
\begin{equation} \label{eqn:CM}
  \{F,G\} = 
    \{f,g\}^{\mathrm{CM}}_{\mm} 
      + \{f,g\}^{\mathrm{CM}}_{\mm,\Bb} + \{f,g\}^{\mathrm{CM}}_{\Bb,\mm}
      + \{f,g\}^{\mathrm{CM}}_{\Bb}
      + \{f,g\}^{\mathrm{CM}}_{n} + \{f,g\}^{\mathrm{CM}}_{s}
\end{equation}
with 
\begin{equation} \label{CM1} 
\begin{system}
  &\{f,g\}^{\mathrm{CM}}_{\mm}
      = 
      - \int \mm \cdot \left(\diff{f}{\mm} \cdot \nabla \diff{g}{\mm}\right)
      + \int \mm \cdot \left(\diff{g}{\mm} \cdot \nabla \diff{f}{\mm}\right) 
  \\
  &\{f,g\}^{\mathrm{CM}}_{\mm, \Bb} 
  = \int \Big(\Bb \times \diff{f}{\mm}\Big) 
      \cdot \Big(\nabla \times \diff{g}{\Bb}\Big)
    - \int \Big(\Bb \times \diff{g}{\mm}\Big) 
    \cdot \Big(\nabla \times \diff{f}{\Bb}\Big)
  \\
  &\{f,g\}^{\mathrm{CM}}_{\Bb}
      = \int \frac{1}{q_i n} \Bb \cdot \left(\Big(\nabla \times \diff{f}{\Bb}\Big) \times \Big(\nabla \times \diff{g}{\Bb} \Big)\right)
  \\
  &\{f,g\}^{\mathrm{CM}}_{n} 
      = 
      - \int n \diff{f}{\mm} \cdot \nabla \diff{g}{n}
      + \int n\diff{g}{\mm} \cdot \nabla \diff{f}{n}
  \\
  &\{f,g\}^{\mathrm{CM}}_{s}
      = 
      - \int s \diff{f}{\mm} \cdot \nabla \diff{g}{s}
      + \int s \diff{g}{\mm} \cdot \nabla \diff{f}{s}
\end{system}
\end{equation}
and as
\begin{equation} \label{eqn:YH}
  \{F, G\} =
  \{\tilde{f}, \tilde{g}\}^{\mathrm{YH}}_{\uu} + \{\tilde{f}, \tilde{g}\}^{\mathrm{YH}}_{\uu, \Bb} + \{\tilde{f}, \tilde{g}\}^{\mathrm{YH}}_{\Bb} + \{\tilde{f}, \tilde{g}\}^{\mathrm{YH}}_{\rho} 
          + \{\tilde{f}, \tilde{g}\}^{\mathrm{YH}}_{s}
\end{equation}
with 
$$
\begin{system}
    \{\tilde{f},\tilde{g}\}^{\mathrm{YH}}_{\uu} 
        &:= \int \frac{1}{\rho} \diff{\tilde{g}}{\uu} \times \diff{\tilde{f}}{\uu} \cdot \nabla \times \uu
    \\
    \{\tilde{f},\tilde{g}\}^{\mathrm{YH}}_{\uu, \Bb} &:= \{f,g\}^{\mathrm{CM}}_{\mm,\Bb}
        = \int \frac{1}{\rho}\Big(\Bb \times \diff{\tilde{f}}{\uu} \Big) \cdot \Big(\nabla \times \diff{\tilde{g}}{\Bb} \Big)
        - \int \frac{1}{\rho}\Big(\Bb \times \diff{\tilde{g}}{\uu} \Big) \cdot \Big(\nabla \times \diff{\tilde{f}}{\Bb} \Big)
    \\
    \{\tilde{f},\tilde{g}\}^{\mathrm{YH}}_{\Bb} &:= \{f,g\}^{\mathrm{CM}}_{\Bb} 
        = \int \frac{m_i}{q_i \rho} \Bb \cdot \Big[\Big(\nabla \times \diff{\tilde{f}}{\Bb}\Big) \times \Big(\nabla \times \diff{\tilde{g}}{\Bb} \Big)\Big]
    \\
    \{\tilde{f},\tilde{g}\}^{\mathrm{YH}}_{\rho}
         &:= \int \diff{\tilde{g}}{\uu} \cdot \nabla \diff{\tilde{f}}{\rho}
          - \int \diff{\tilde{f}}{\uu} \cdot \nabla \diff{\tilde{g}}{\rho}
    \\
    \{\tilde{f},\tilde{g}\}^{\mathrm{YH}}_{s} &:= \{f,g\}^{\mathrm{CM}}_{s} 
         = \int \frac{s}{\rho} \diff{\tilde{g}}{\uu} \cdot \nabla \diff{\tilde{f}}{s}
          - \int \frac{s}{\rho} \diff{\tilde{f}}{\uu} \cdot \nabla \diff{\tilde{g}}{s}.
\end{system}
$$
By inspection of the brackets appearing in the right-hand sides of 
\eqref{eqn:CM} and \eqref{eqn:YH} we obtain the following result.
\begin{theorem}
  The Hall-MHD bracket 
  \eqref{eqn:brackets_hmhd} is equivalent
  with the ones proposed in \cite{yoshida_canonical_2013}
  and \cite{coquinot2020general}.
\end{theorem}

\section{Viscoresistive Hall MHD}
\label{sec:vrhmhd}

\subsection{Lagrangian Variational approach}
We now use the formalism of \cite{gay2017lagrangian1,gay2017lagrangian2} to add dissipation to our variational model and recover viscoresistive Hall-MHD. Specifically, this amounts in adding viscous and resistive terms 
in the constrained entropy variations, which involve respectively a stress tensor $\sigma$ of order 2 and a scalar resistivity $\eta \ge 0$.
Here the viscous term is incorporated exactly as in these works,
while for the resistive part we add a force corresponding to the interpretation of Drude's model, which depends on the difference between both velocities. This models a resistive force that will draw the electron flow closer to the ion flow.
\begin{theorem}
Consider the following lagrangian:
\begin{subequations}
\label{eqn:lagrangians_vrhmhd}
\begin{equation}
\label{eqn:lagrangian_vrhmhd}
l(\uu_i, n, s, \uu_e, \Ab) = \int_\Omega \frac{m_i n |\uu_i|^2}{2} - n e(n, s) + q_i n (\uu_i-\uu_e) \cdot \Ab - \frac{|\nabla \times \Ab|^2}{2 \mu_0}.
\end{equation}
\end{subequations}
Then extremizers of the action $\mathfrak{S} = \int_0^T l dt$ under constrained variations 
\begin{align} 
  \label{eqn:vrhmhd_constraints}
    \delta \uu_i = \partial_t \vv_i + [\uu_i,\vv_i], 
    \qquad 
    \delta n = -\Div (n \vv_i), \\
    \frac{\delta l}{\delta s}(\delta s +\Div (s \vv_i)) = -\sigma : \nabla \vv_i 
      - \eta (q n)^2 (\uu_i-\uu_e) \cdot (\vv_i-\vv_e) \\
    \delta \uu_e = \partial_t \vv_e + [\uu_e,\vv_e],  
\end{align}
with arbitrary time-dependent vector fiels $\vv_i$ and $\vv_e$ vanishing at $t=0$ and $T$, %
supplemented with the advection equations:
\begin{subequations}
\begin{equation}
\label{eqn:advection_rhoi_vrhmhd}
\partial_t n + \Div(n \uu_i) = 0 ~,
\end{equation}
\begin{equation}
\label{eqn:advection_s_vrhmhd}
\frac{\delta l}{\delta s}(\partial_t s + \Div(s \uu_i)) = -\sigma : \nabla \uu_i - \eta (q_in)^2 \abs{\uu_i-\uu_e}^2~,
\end{equation}
\end{subequations}
are solution to the following system of equations:
\begin{subequations}
\label{eqn:vrhmhd}
\begin{equation}
\label{eqn:vrhmhd_massi}
\partial_t n + \Div(n \uu_i) = 0 ~,
\end{equation}
\begin{equation}
\label{eqn:vrhmhd_s}
\frac{\delta l}{\delta s}(\partial_t s + \Div(s \uu_i)) = -\sigma : \nabla \uu_i - \eta \abs{\Jb}^2 ~.
\end{equation}
\begin{equation}
\label{eqn:vrhmhd_momi}
m_i n ( \partial_t \uu_i + \uu_i \cdot \nabla \uu_i) + \nabla p - q_in (\Eb + \uu_i \times \Bb) 
  - \nabla \cdot \sigma + \eta q_i n \Jb  = 0 ~, 
\end{equation}
\begin{equation}
\label{eqn:vrhmhd_mome}
\Eb + \uu_e \times \Bb = \eta \Jb ~, 
\end{equation}
\begin{equation}
\label{eqn:vrhmhd_Ampere}
\frac{\nabla \times \Bb}{\mu_0} = \Jb~,
\end{equation}
\end{subequations}
where $\Jb := q_i n (\uu_i - \uu_e)$ is the current density.
\end{theorem}
\begin{proof}
\cref{eqn:vrhmhd_massi,eqn:vrhmhd_s} are the advection equation imposed and \cref{eqn:vrhmhd_Ampere} are obtained exaclty as in the proof of \cref{thm:shmhd}. Now looking at all the contributions in $\vv_i$ and $\vv_e$ 
\begin{align*}
0 &= \int_0^T \int_\Omega \frac{\delta l}{\delta \uu_i} \delta \uu_i + \frac{\delta l}{\delta n} \delta n + \frac{\delta l}{\delta s} \delta s + \frac{\delta l}{\delta \uu_e} \delta \uu_e \\
  &= \int_0^T \int_\Omega \frac{\delta l}{\delta \uu_i} (\partial_t \vv_i + [\uu_i, \vv_i]) - \frac{\delta l}{\delta n} \nabla \cdot (n \vv_i) - \frac{\delta l}{\delta s} \nabla \cdot (s \vv_i) -\sigma : \nabla \vv_i 
  - \eta (q_i n)^2 (\uu_i-\uu_e) \cdot \vv_i \\
  &+ \eta (q_i n)^2 (\uu_i-\uu_e) \cdot  \vv_e + \frac{\delta l}{\delta \uu_e} (\partial_t \vv_e + [\uu_e, \vv_e])~,
\end{align*}
considering variations only in $\vv_i$ and using the same computation as before (as we only added one term) : 
\begin{align*}
0 &= \int_0^T \int_\Omega \frac{\delta l}{\delta \uu_i} (\partial_t \vv_i + [\uu_i, \vv_i]) - \frac{\delta l}{\delta n} \nabla \cdot (n \vv_i) - \frac{\delta l}{\delta s} \nabla \cdot (s \vv_i) -\sigma : \nabla \vv_i - \eta (q_i n)^2 (\uu_i-\uu_e) \cdot \vv_i \\
  &= \int_0^T \int_\Omega -\big(m_i n( \partial_t \uu_i + \uu_i \cdot \nabla \uu_i) + \nabla p - q_i n (\Eb + \uu_i \times \Bb)\big) \cdot \vv_i + \nabla \cdot \sigma \cdot \vv_i - \eta (q_i n)^2 (\uu_i-\uu_e) \cdot \vv_i.
\end{align*}
This being true for all $\vv_i$, we find \cref{eqn:vrhmhd_momi}.
Considering next variations in $\vv_e$ and computing 
as in the proof of \cref{thm:shmhd},
we write
\begin{align*}
0 &= \int_0^T \int_\Omega \eta q_i n \Jb \cdot \vv_e + \frac{\delta l}{\delta \uu_e} (\partial_t \vv_e + [\uu_e, \vv_e])\\
  &= \int_0^T \int_\Omega q_i n \big(\eta \Jb - (\Eb + \uu_e \times \Bb)\big) \cdot \vv_e
\end{align*}
which gives \cref{eqn:vrhmhd_mome}.
\end{proof}
To recover the viscoresistive Hall-MHD equations from \eqref{eqn:vrhmhd} we next proceed as follows: summing \cref{eqn:vrhmhd_momi,eqn:vrhmhd_mome}, and 
using the definition of the current
($\Jb = q_i n (\uu_i - \uu_e)$), we find
\begin{equation}
\label{eqn:vrhmhd_momJ}
m_i n ( \partial_t \uu_i +  \uu_i \cdot \nabla \uu_i) + \nabla p - \Jb\times \Bb - \nabla \cdot \sigma = 0 ~.
\end{equation}
Ohm's law is obtained by combining \cref{eqn:vrhmhd_mome} with the definition of the current,
\begin{equation}
\label{eqn:vrhmhd_OhmJ}
\Eb + \uu_i \times \Bb = \frac{1}{q_i n}  \Jb \times \Bb +  \eta \Jb ~,
\end{equation}
and from the definition of the fields we recover Faraday's equation:
\begin{equation}
\label{eqn:vrhmhd_Faraday}
\partial_t \Bb = - \nabla \times \Eb 
  = -\nabla \times \Big(\Big(\frac{1}{q_i n} \Jb - \uu_i\Big) \times \Bb + \eta \Jb \Big)~.
\end{equation}

\subsection{Metriplectic bracket, Hamiltonian and Entropy for Viscoresistive Hall MHD}

We finally turn to the metriplectic reformulation of the previous equations. 
The conservative bracket and Hamiltonian will remain the same as in the non-dissipative case of the previous section, while we will now consider an entropy (being simply the total entropy of the system) and a dissipative bracket, that will enable us to take into account the two dissipative mechanisms we incorporated in our equations. 
We use the formalism introduced in \cite{morrison2024inclusive} where the dissipation is included using a 4-bracket that have the same symmetries as a curvature tensor. 
This bracket take as argument twice the Hamiltonian on the second and fourth slot, and an entropy to be dissipated on the third position. The first slot is as in the standard Poisson bracket/Hamiltonian formulation occupied by the function whose dynamic is described.
We make the assumption that the viscous tensor has the form 
\begin{equation}
\label{eqn:structure_viscosity}
\sigma = \Lambda \nabla \uu_i ~,
\end{equation}
where $\Lambda$ is a 4-tensor that has the symmetry 
$\Lambda_{abcd} = \Lambda_{cdab}$ and is positive 
in the sense that 
$v : \Lambda v = \sum_{abcd} v_{ab} \Lambda_{abcd} v_{cd} \geq 0$ 
for any 2-tensor $v$.

\begin{theorem}
Let $H = H(\mm_i, \mm_e, n, s)$ and $\{\cdot, \cdot\}$ 
be the Hamiltonian 
and bracket from \cref{thm:hmhd_symplectic}.
A curve $t \to (\uu, n, s, \Ab)(t)$ is a solution to the equations 
\cref{thm:shmhd} iff 
the corresponding curve 
$t \to (\mm_i, \mm_e, n, s)(t)$ 
solves the metriplectic system 
\begin{equation}
  \label{eqn:vrhmhd_metriplectic}
  \dot{F} = \{F, H\} + (F,H;S,H)
\end{equation}
with the total entropy functional 
\begin{equation}
\label{eqn:hmhd_tot_entropy}
S= \int_\Omega s ~,
\end{equation}
and a disipative bracket defined by
\begin{subequations} \label{eqn:hmhd_dissip_bracket}
\begin{equation}  
(F,M;G,N)=(F,M;G,N)_{visc}+(F,M;G,N)_{res} ~,
\end{equation}
with 
\begin{equation}
  \label{eqn:hmhd_visc_bracket}
  (F,M;G,N)_{visc} = \int_{\Omega}\frac{1}{T} \left(
    \Big(
      \frac{\delta F}{\delta s}
        \nabla\frac{\delta M}{\delta \mm_i} 
        - 
        \frac{\delta M}{\delta s}
        \nabla\frac{\delta F}{\delta \mm_i} 
        \Big)
        : \Lambda
        \Big(
          \frac{\delta G}{\delta s}
            \nabla\frac{\delta N}{\delta \mm_i} 
            - 
            \frac{\delta N}{\delta s}
            \nabla\frac{\delta G}{\delta \mm_i} 
            \Big)        
      \right)
\end{equation}
and
\begin{equation}
  \label{eqn:hmhd_res_bracket}
  \begin{aligned}
    (F,M;G,N)_{res} = \int_{\Omega}\frac{\eta (q_i n)^2}{T} 
      &\left(
      \Big(
      \frac{\delta F}{\delta s} 
      \Big(
        \frac{\delta M}{\delta \mm_i}-\frac{\delta M}{\delta \mm_e}
      \Big)
      - 
      \frac{\delta M}{\delta s} 
      \Big(
        \frac{\delta F}{\delta \mm_i}-\frac{\delta F}{\delta \mm_e}
      \Big)
      \Big)
      \right.
      \\
      &\left.
        \quad \cdot 
      \Big(
        \frac{\delta G}{\delta s} 
        \Big(
          \frac{\delta N}{\delta \mm_i}-\frac{\delta N}{\delta \mm_e}
        \Big)
        - 
        \frac{\delta N}{\delta s} 
        \Big(
          \frac{\delta G}{\delta \mm_i}-\frac{\delta G}{\delta \mm_e}
        \Big)
      \Big)
  \right)
  \end{aligned}
\end{equation}
\end{subequations}
where the temperature is defined as 
$
T = \frac{\delta H}{\delta s}.
$
\end{theorem}

\begin{remark}
The above definitions meet the 
requirements of metriplectic dynamics
\cite{morrison2024inclusive}:
the entropy functional is a Casimir
($\{F,S\} = 0$ for any $F$),
and the metric 4-bracket has the proper symmetries, namely
$(F,M;G,N) = - (M,F;G,N) = - (F,M;N,G) = (G,N;F,M)$ 
hold for any functionals $F,G,M,N$.
\end{remark}

\begin{remark}
  Using
  $\frac{\delta H}{\delta s} = T$,
  $\frac{\delta H}{\delta \mm_i} = \uu_i$
  and 
  $\frac{\delta H}{\delta \mm_e} = \uu_e$,
  we find that the symmetric 2-bracket \cite{morrison2024inclusive}
  corresponding to \eqref{eqn:hmhd_dissip_bracket} reads 
  \begin{subequations}
    \label{eqn:hmhd_dissip_bracket_2}
  \begin{equation}
    (F,G)_H = (F,G)_{H, visc} + (F,G)_{H, res} 
  \end{equation}
  with 
  \begin{equation}
    \label{eqn:hmhd_visc_bracket_2}
    (F,G)_{H, visc} = (F,H;G,H)_{visc} 
      = \int_\Omega \frac{1}{T}
    \left(
        \Big(
          \frac{\delta F}{\delta s}
            \nabla \uu_i
            - 
            T \nabla\frac{\delta F}{\delta \mm_i} 
            \Big)
            : \Lambda
            \Big(
              \frac{\delta G}{\delta s}
                \nabla\uu_i
                - 
                T \nabla\frac{\delta G}{\delta \mm_i} 
                \Big)        
          \right)      
  \end{equation}
  and 
  \begin{equation}
    \label{eqn:hmhd_res_bracket_2}
    \begin{aligned}
      (F,G)_{H, res} = (F,H;G,H)_{res} 
      = \int_{\Omega}\frac{\eta (q_i n)^2}{T} 
        &\left(
        \Big(
        \frac{\delta F}{\delta s} 
        (\uu_i-\uu_e)
        - 
        T
        \Big(
          \frac{\delta F}{\delta \mm_i}-\frac{\delta F}{\delta \mm_e}
        \Big)
        \Big)
        \right.
        \\
        &\left.
          \quad \cdot 
        \Big(
          \frac{\delta G}{\delta s} 
          (\uu_i-\uu_e)
          - 
          T
          \Big(
            \frac{\delta G}{\delta \mm_i}-\frac{\delta G}{\delta \mm_e}
          \Big)
        \Big)
    \right).
    \end{aligned}
  \end{equation}
  \end{subequations}
\end{remark}

We may now prove the theorem.
\begin{proof}
Using $\frac{\delta S}{\delta \mm_i} = 0$ and $\frac{\delta S}{\delta s}=1$, we derive from 
\eqref{eqn:hmhd_visc_bracket_2} and \eqref{eqn:hmhd_res_bracket_2} that
$$ 
(F,H;S,H)_{visc}
  = \int_\Omega \frac{1}{T}
  \left(
    \Big(
      \frac{\delta F}{\delta s}
        \nabla \uu_i
        - 
        T \nabla\frac{\delta F}{\delta \mm_i} 
        \Big)
        : \Lambda
            \nabla\uu_i
      \right)      
  = \int_{\Omega} \frac{1}{T} \nabla \uu_i : \sigma \frac{\delta F}{\delta s} 
  + \nabla \cdot \sigma \cdot \frac{\delta F}{\delta \mm_i}
$$ 
where we remind that $\sigma$ is given by \eqref{eqn:structure_viscosity},
and 
$$ 
(F,H;S,H)_{res} 
    = \int_\Omega \eta (q_i n)^2 \left(\frac{\abs{\uu_i-\uu_e}^2}{T} \frac{\delta F}{\delta s}
			  - (\uu_i-\uu_e) \cdot \Big(\frac{\delta F}{\delta \mm_i}-\frac{\delta F}{\delta \mm_e}\Big).
        \right)
$$ 
Doing the same computations as in the proof of \cref{thm:hmhd_symplectic} and using the definitions, one then finds
\begin{equation*}
\dot{F} 
= \int_\Omega \frac{\delta F}{\delta \mm_i} \dot{\mm_i} + \frac{\delta F}{\delta \mm_e} \dot{\mm_e} + \frac{\delta F}{\delta n} \dot{n} + \frac{\delta F}{\delta s} \dot{s} 
		= \{F,H\} + (F,H;S,H).
\end{equation*}
\end{proof}

Plugging $S$ in the equation above we also recover the total entropy variation (that could already be obtained from \cref{eqn:vrhmhd_s} )
\begin{equation}
\label{eqn:hmhd_entrop_dissip}
\frac{dS}{dt}= \int_\Omega  \frac{\nabla \uu_i : \Lambda \nabla \uu_i}{T} + \frac{\eta (q_i n)^2 \abs{\uu_i-\uu_e}^2}{T}
\end{equation}
which is positive due to our assumption on $\Lambda$ and the fact that $\eta \ge 0$. 

\subsection{Relation with the metric bracket of Coquinot-Morrison}

We reformulate our 2-bracket \eqref{eqn:hmhd_dissip_bracket_2}
in $y = (\mm, \Bb, n, s)$ variables. 
Writing again $f(y) = F(Y)$,
we remind that \eqref{eqn:chgvar} gives 
$\frac{\delta F}{\delta \mm_i} = \frac{\delta f}{\delta \mm}$,
$\frac{\delta F}{\delta s} = \frac{\delta f}{\delta s}$,
and 
$\frac{\delta F}{\delta \mm_i}-\frac{\delta F}{\delta \mm_e}
= \frac{1}{q_i n} \nabla \times \diff{f}{\Bb}$.
Together with $\uu_i - \uu_e = \frac{1}{q_i n} \Jb$, this yields
\begin{subequations}
  \label{eqn:hmhd_dissip_bracket_CM}
\begin{equation}
  (F,G)_H = (f,g)^{CM} = (f,g)^{CM}_{visc} + (f,g)^{CM}_{res} 
\end{equation}
with 
\begin{equation}
  \label{eqn:hmhd_visc_bracket_CM}
(f,g)^{CM}_{visc} = \int_\Omega T 
    \Big(
      \frac{1}{T} \frac{\delta f}{\delta s}
      \nabla \uu_i
      - 
      \nabla\frac{\delta f}{\delta \mm} 
      \Big)
      : \Lambda
      \Big(
        \frac{1}{T} \frac{\delta g}{\delta s}
          \nabla\uu_i
          - 
          \nabla\frac{\delta g}{\delta \mm} 
          \Big)         
\end{equation}
and 
\begin{equation}
  \label{eqn:hmhd_res_bracket_CM}
  \begin{aligned}
    (f,g)^{CM}_{res}
    \int_{\Omega} T 
    &
    \Big(
      \frac{1}{T}\frac{\delta f}{\delta s} 
    \Jb
    - 
    \nabla \times \diff{f}{\Bb}
    \Big)
    \cdot \eta
    \Big(
      \frac{1}{T} \frac{\delta g}{\delta s} 
      \Jb
      - 
      \nabla \times \diff{g}{\Bb}
    \Big)
  \end{aligned}
\end{equation}
\end{subequations}
which corresponds to the metric bracket 
(4.23) of \cite{coquinot2020general}, with 
$$
\begin{system}
  &\Lambda_{vv} = \Lambda 
  \\
  &\Lambda_{vj} = \Lambda_{jv} = \Lambda_{jj} = 0
  \\  
\end{system}
\qquad \text{ and } \qquad 
\begin{system}
  &\eta_{jj} = \eta 
  \\
  &\eta_{TT} = \eta_{Tj} = \eta_{jT} = 0 ~.
\end{system}
$$

\section{Conclusion}
In this work we presented variational principles for models of Hall MHD and extended MHD, using simplification from a Lagrangian describing the evolution of two fluids in an electromagnetic fluid.
Viscosity and resistivity where added using a generalized Lagrange-D'Alembert principle, and all the least action principles were reformulated in the metriplectic framework.
Those least-action principle were then reformulated in the metriplectic framework, using a canonical Legendre transform and standard Lie-Poisson bracket for the symplectic part.

Future work will focus on combining this work and the variational numerical scheme described by the authors in \cite{carlier2025variational} to build new approaches to the numerical resolution of Hall and extended MHD.
\label{sec:conclusion}
\bibliographystyle{plain} 
\bibliography{var_MHD}
\end{document}